\documentclass[10pt]{article}

\usepackage[letterpaper]{geometry}
\usepackage{HICSS_Latex_Template/hicss}
\usepackage{times}
\usepackage{url}
\usepackage{latexsym}
\usepackage{indentfirst}
\usepackage{graphicx}
\usepackage{amsmath}
\usepackage{amssymb}
\usepackage{booktabs}
\usepackage{multirow}
\usepackage{enumitem}
\usepackage{float}
\usepackage[section]{placeins}
\usepackage[style=apa]{biblatex}

\AtBeginBibliography{\small}

\title{Toward Sustainable AI Deployment: A Carbon-Aware Decision Framework for Enterprise Supply Chain Systems}

\author{
\begin{tabular}{c@{\hspace{2em}}c@{\hspace{2em}}c}
Haoran Yu\textsuperscript{*} & Lifei Liu & Danping Zhang \\
University of Florida & Wichita State University & Nanchang Hangkong University \\
United States & United States & China \\
\texttt{haoranyu889@gmail.com} & \texttt{lliu.lifei@gmail.com} & \texttt{zhangdanping@nchu.edu.cn}
\end{tabular}
}

\date{}

\begin{document}
\maketitle
{\renewcommand{\thefootnote}{\fnsymbol{footnote}}%
\footnotetext[1]{Corresponding author.}}

\begin{abstract}
Enterprises deploying AI for supply chain decisions commonly default to the largest available language model, a procurement heuristic that neglects both empirical performance and environmental cost. We benchmark six large language models across 520 supply chain tasks, simultaneously measuring decision quality and estimated generation-related operational carbon. Drawing on the Technology-Organization-Environment (TOE) framework, we develop a Carbon-Aware AI Procurement Framework (CAAPF), a Green IS design artifact that operationalizes sustainable AI governance for enterprise procurement. Within this bounded sample, quality spans 0.497--0.723, and the models with the largest disclosed parameter totals do not achieve the highest scores. The design does not isolate size, provider, architecture, or benchmark-construction effects. A category-by-tier calibrated GreenRoute proof of concept reaches 0.733 mean out-of-sample quality at an estimated 0.402~gCO\textsubscript{2}/task. Static Haiku reaches 0.699 at 0.022~gCO\textsubscript{2}/task, while Sonnet reaches 0.723 at 0.401~gCO\textsubscript{2}/task, demonstrating that the preferred strategy depends on the organization's quality requirement. Our ``benchmark first, select green'' principle suggests that environmental responsibility and decision quality can be mutually reinforcing, contributing to sustainable digital infrastructure governance aligned with SDG~12 and SDG~13.

\end{abstract}

\section{Introduction}

Recent work examines language models as interfaces to supply chain decision tools and optimization systems \parencite{simchilevi2025llm, li2023llmsc, huang2026dacri}. Model procurement often defaults to the largest available language model on the assumption that more parameters yield better decisions.

This size-based heuristic can carry substantial environmental cost. LLM inference produces carbon dioxide emissions through the electricity consumed by GPU clusters \parencite{luccioni2024power, wiesner2025efficiency}. Across repeated inventory, supplier-risk, and forecasting queries, differences in per-task energy use can accumulate into a material operational footprint. The United Nations Sustainable Development Goals (SDGs) 12 and 13 call for responsible consumption and climate action \parencite{sdg2030}, but environmental impact is not routinely integrated into model-selection decisions.

This paper addresses the intersection of Green Information Systems (Green IS), supply chain decision-making, and AI model selection. Prior work has quantified training-time carbon costs \parencite{strubell2019energy, patterson2022carbon} and proposed cost-efficient routing \parencite{chen2023frugalgpt, ong2024routellm}. Evidence remains limited on how quality and operational carbon vary together on supply-chain-specific tasks and on how organizations should translate such measurements into an auditable procurement decision.

We address three research questions:

\begin{description}[leftmargin=1.2cm,labelwidth=1cm,labelsep=0.2cm,topsep=2pt,itemsep=2pt]
\item[RQ1:] How does the quality--carbon tradeoff vary across LLMs for supply chain decision tasks?
\item[RQ2:] How are observed differences in model size, model family, and architecture associated with decision quality in the evaluated supply chain tasks?
\item[RQ3:] How can a theoretically grounded framework guide threshold-contingent, sustainable AI model selection?
\end{description}

Drawing on TOE \parencite{tornatzky1990processes}, we develop and empirically examine a Carbon-Aware AI Procurement Framework (CAAPF). Our contributions are fourfold:

\begin{enumerate}[leftmargin=*]
  \item A \textbf{threshold-contingent procurement framework} that links domain benchmarking to organization-specific quality, carbon, cost, risk, and governance constraints.

  \item \textbf{Descriptive evidence} that nominal model size is an unreliable procurement proxy in the six-model sample, together with explicit limits on cross-architecture, provider, and benchmark-style inference.

  \item Identification of a \textbf{model-specific deliberation pattern} in which longer DeepSeek R1 responses coincide with lower accuracy and higher per-task carbon on formula tasks.

  \item A \textbf{boundary-aware evaluation} showing when simple model selection or escalation is preferable to routing and when task-level routing may add value.
\end{enumerate}

\section{Theoretical Background}

\subsection{Green IS and Sustainable AI}

Green Information Systems research examines how IT artifacts can be designed, deployed, and governed to enable environmentally responsible outcomes \parencite{melville2010sustainability}. Within this tradition, the environmental impacts of AI systems represent an emerging concern: while AI can support sustainability goals, the computational infrastructure underlying AI systems itself carries significant environmental costs \parencite{wu2022sustainable, verdecchia2023systematic}.

The concept of Green AI was formalized by \textcite{schwartz2020green}, who argued that the AI community disproportionately rewards accuracy gains achieved through computational brute force while neglecting efficiency. \textcite{strubell2019energy} showed that extensive architecture search and tuning can dominate the emissions of a final training run. \textcite{luccioni2024power} extended measurement to inference and found large energy differences across architectures and task categories. \textcite{wiesner2025efficiency} argued that growing inference-time compute for reasoning models can outpace hardware-efficiency gains.

On the measurement side, \textcite{anthony2020carbontracker} developed Carbontracker for real-time energy monitoring, while \textcite{lacoste2019quantifying} created the Machine Learning Emissions Calculator. \textcite{dodge2022measuring} demonstrated that carbon intensity varies substantially with cloud region and time of day, suggesting scheduling and geographic placement as mitigation levers. \textcite{samsi2023words} benchmarked LLM inference energy costs and noted that inference energy had received less attention than training energy.

For supply chain applications, \textcite{simchilevi2025llm} described LLM interfaces for explaining tool recommendations, exploring what-if scenarios, and updating decision models. \textcite{li2023llmsc} combined an LLM interface with optimization code and identified ambiguity, generated-code errors, and out-of-distribution use as limitations. Moving from explanation toward prescriptive decision support, \textcite{huang2026dacri} developed a decision-aware causal intervention ranking approach for critical supply chains, prioritizing interventions by their estimated causal effect on outcomes rather than by predictive fit alone. These concerns motivate direct evaluation of numerical and domain-specific decisions rather than assuming that model scale will transfer into operational quality.

\subsection{Model Routing and Selection}

The economics of LLM deployment have motivated research into intelligent model selection. \textcite{chen2023frugalgpt} introduced FrugalGPT, reducing inference cost by up to 98\% through LLM cascades that route easy queries to cheap models and hard queries to expensive ones. \textcite{ong2024routellm} developed RouteLLM, learning routing policies from preference data with substantial cost savings and minimal quality degradation. \textcite{cruciani2025choosing} explicitly connected model selection to environmental sustainability, calling for empirical validation across specific domains. \textcite{sardana2024beyond} extended scaling laws to account for inference compute, showing that optimal model size depends on query volume. Neither work included supply chain tasks or carbon measurements.

\subsection{TOE as Analytical Framework for AI Procurement}

TOE explains organizational technology adoption through three contexts: the \emph{technology} context covers the attributes and availability of candidate technologies; the \emph{organization} context covers readiness, resources, governance, and operational requirements; and the \emph{environment} context covers competition, regulation, and external stakeholder pressure \parencite{tornatzky1990processes, baker2012toe, zhu2006assimilation}. Green IS research further treats information-system choices as mechanisms through which organizations can act on environmental objectives \parencite{melville2010sustainability}. We use TOE to specify what an empirical benchmark alone cannot determine.

First, the \textbf{Technology context} directs attention to relevant attributes of candidate technologies; we operationalize these attributes through observed task quality and resource intensity rather than a parameter-count proxy. Second, the \textbf{Organization context} identifies the internal conditions that determine admissibility: decision criticality, an acceptable-quality threshold ($\tau$), API cost, latency, data governance, vendor risk, and oversight requirements. Third, the \textbf{Environment context} identifies external sustainability and accountability pressures that motivate emissions documentation and periodic re-evaluation. CAAPF translates these contexts into decision rules rather than using TOE only to label benchmark variables.

We therefore derive three propositions. P1 and P2 are examined empirically, whereas P3 is instantiated as a design proposition whose organizational effects require later validation. \textit{P1: Nominal model size does not reliably rank technology relative advantage when quality and estimated generation-related carbon are evaluated jointly.} \textit{P2: The preferred selection strategy changes with organization-specific quality and risk thresholds, so simple selection may dominate routing at moderate thresholds.} \textit{P3: A TOE-based process produces an auditable model-selection record that incorporates environmental pressure without treating carbon as the only procurement criterion.}

\section{Carbon-Aware AI Procurement Framework}

\begin{figure}[t]
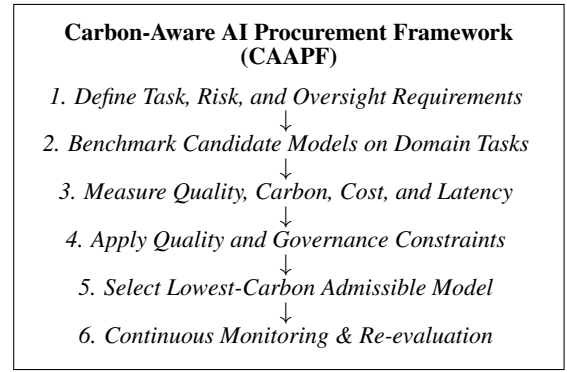

\centering
\fbox{\parbox{0.9\columnwidth}{\small\centering
\vspace{4pt}
\textbf{Carbon-Aware AI Procurement Framework (CAAPF)}\\[6pt]
\begin{tabular}{c}
\textit{1. Define Task, Risk, and Oversight Requirements} \\
$\downarrow$ \\
\textit{2. Benchmark Candidate Models on Domain Tasks} \\
$\downarrow$ \\
\textit{3. Measure Quality, Carbon, Cost, and Latency} \\
$\downarrow$ \\
\textit{4. Apply Quality and Governance Constraints} \\
$\downarrow$ \\
\textit{5. Select Lowest-Carbon Admissible Model} \\
$\downarrow$ \\
\textit{6. Continuous Monitoring \& Re-evaluation} \\
\end{tabular}
\vspace{4pt}
}}
\caption{CAAPF links TOE contexts to an auditable procurement sequence. Technology is profiled in Steps 2--3, organizational constraints define admissibility in Steps 1 and 4, and environmental accountability motivates monitoring.}
\label{fig:caapf}
\end{figure}

Figure~\ref{fig:caapf} presents CAAPF, a decision process for sustainable AI procurement. It separates the organization's admissibility decision from the technical optimization performed after admissible candidates have been identified.

The framework's core principle is \textbf{``benchmark first, select green''}. The selection process is: define a representative task portfolio; set $\tau$ from the operational consequences of error; record non-quality constraints such as API cost, latency, security, data residency, and vendor risk; benchmark every candidate under a common prompt protocol; form the set satisfying all constraints; and select the lowest-carbon member. CAAPF can apply quality requirements at the portfolio level or, for heterogeneous workloads, separately to task classes. A model that passes an aggregate threshold is not necessarily admissible when individual task classes must each meet a service floor. A router is justified only if task-level heterogeneity improves this rule relative to simple selection or escalation. Monitoring reopens the decision when models, workloads, prices, or grid conditions change.

We position CAAPF as a \emph{Green IS design artifact} that makes model selection inspectable. Table~\ref{tab:decision_matrix} states the general decision logic without transferring numerical cutoffs from this benchmark to other organizations.

\begin{table}[t]
\caption{CAAPF decision matrix; thresholds must be calibrated to organizational risk.}
\label{tab:decision_matrix}
\centering
\scriptsize
\begin{tabular}{@{}p{1.75cm}p{2.2cm}p{2.35cm}@{}}
\toprule
\textbf{Requirement} & \textbf{Strategy} & \textbf{Decision Rule} \\
\midrule
Moderate portfolio quality & Static low-carbon model & Select the lowest-carbon single model satisfying constraints \\
Higher quality met by one model & Lowest-carbon passing model & Routing adds no value if one model passes all constraints \\
Uneven quality across task classes & Calibrated routing & Route only if a mixture improves the feasible frontier \\
Safety-critical & Human or deterministic oversight & Require validation and an audit trail \\
\bottomrule
\end{tabular}
\end{table}

In practice, organizations should derive $\tau$ from task-level loss, regulatory duties, existing human or system performance, and required oversight, and should reject deployment when no tested configuration satisfies those conditions. In our GreenRoute implementation, the study-specific value $\tau=0.65$ is applied separately to each of 18 category$\times$tier calibration cells, not only to aggregate portfolio quality. Thus, Haiku's aggregate score of 0.699 does not by itself establish that it passes every task-class requirement.

\section{Methodology}

\subsection{Task Taxonomy}

We construct a benchmark of 520 supply chain tasks spanning six operational categories: Demand Forecasting (92 tasks; 33 using real M5 Competition data \parencite{makridakis2022m5}, a setting where large-scale retail forecasting must balance accuracy against operational stability \parencite{li2026stability}), Vehicle Routing (84), Inventory Optimization (96), Supplier Risk Assessment (88), Order Fulfillment (80), and Demand Classification (80).

Three difficulty tiers reflect decision complexity: \textbf{Tier~1} (formula application, $\sim$35\%): single-step calculations with deterministic solutions (e.g., computing EOQ). \textbf{Tier~2} (multi-factor reasoning, $\sim$40\%): problems requiring integration of multiple inputs and conditional logic. \textbf{Tier~3} (strategic judgment, $\sim$25\%): open-ended decisions requiring synthesis of quantitative analysis with qualitative factors.

Tasks were authored by experts in operations management following a structured protocol based on established supply chain planning topics \parencite{chopra2019supply}. For each category $\times$ tier combination, tasks span recurrent forecasting, inventory, routing, risk, fulfillment, and classification problem families while maintaining consistent difficulty calibration.

Each task underwent three-stage validation: (a) independent solution verification by a separate author, (b) full-team review for domain accuracy and difficulty calibration, and (c) pilot testing on two models (one small, one large) to confirm meaningful performance variation. We acknowledge that tasks were not validated by external practicing supply chain professionals, and real-world decisions involve richer organizational context that single-prompt evaluation does not capture.

\subsection{Supply Chain Domain Knowledge Tests}

Beyond general task performance, we design 30 ``SC-trap'' tasks that specifically probe supply chain domain knowledge. These tasks are constructed so that a model lacking domain expertise produces plausible-sounding but incorrect answers. Examples include testing the $\sqrt{\text{LT}}$ safety stock relationship, bullwhip quantification, risk pooling benefits, EOQ sensitivity, and newsvendor critical ratio application.

\subsection{Models and Carbon Estimation}

Table~\ref{tab:models} presents the six LLMs evaluated. The availability-based sample covers four providers, but four of the six models come from Anthropic and Meta; it is not representative of the broader LLM market. Provider disclosures report both total and active parameters for several MoE models \parencite{meta2025llama4, mistral2025large3, deepseek2024v3}, while Anthropic does not disclose parameter counts for the evaluated Claude models. Dense totals, MoE totals, and active parameters are not equivalent measures of inference compute, so they are reported descriptively and are not used in a cross-model correlation. Following prior measurement work \parencite{luccioni2024power, dodge2022measuring}, Table~\ref{tab:models}'s gCO\textsubscript{2}/Mtok coefficients are study-specific engineering estimates based on assumed serving hardware, GPU power, throughput, PUE, and grid intensity, not provider measurements or direct datacenter observations. Study assumptions are PUE 1.1--1.2 and US-East intensity 0.38~kgCO\textsubscript{2}/kWh; all estimates carry $\pm$50\% uncertainty. Generation-related per-task carbon is calculated as model-specific gCO\textsubscript{2}/Mtok $\times$ mean output tokens $/10^{6}$. Because complete input/prefill token records were not retained, these values estimate output-generation carbon rather than complete end-to-end serving energy.

\begin{table}[t]
\caption{Models evaluated. Carbon estimates carry $\pm$50\% uncertainty.}
\label{tab:models}
\centering
\scriptsize
\begin{tabular}{@{}llp{2.25cm}r@{}}
\toprule
\textbf{Model} & \textbf{Provider} & \textbf{Reported Parameters} & \textbf{gCO\textsubscript{2}/Mtok} \\
\midrule
Claude Sonnet 4.6  & Anthropic  & Not disclosed & 900  \\
Claude Haiku 4.5   & Anthropic  & Not disclosed & 70   \\
Mistral Large 3    & Mistral    & 675B total / 41B active & 5,500 \\
Llama 4 Scout      & Meta       & 109B total / 17B active & 60   \\
Llama 3.3 70B      & Meta       & 70B dense total & 980  \\
DeepSeek R1        & DeepSeek   & 671B total / 37B active & 6,000 \\
\bottomrule
\end{tabular}\\[2pt]
{\scriptsize Parameter definitions are architecture-specific and not directly comparable.}
\end{table}

\subsection{Prompting and Evaluation Framework}

All candidate models receive the same task text and category-specific system instruction. Task-specific output caps are 1,024, 1,536, or 2,048 tokens, and candidate-model temperature is set to zero. This fixed protocol controls prompt wording across models but does not test how alternative prompting, demonstrations, or output constraints affect quality, response length, and carbon.

We employ LLM-as-Judge evaluation \parencite{zheng2023judging} with two independent judges from different providers: Claude Opus~4 (Anthropic) and Qwen3-235B (Alibaba). Each judge scores numeric accuracy, reasoning quality, and SC domain knowledge on 1--10 scales; the three dimensions are averaged equally and then normalized. Qwen uses temperature zero, while Opus uses its provider default because the endpoint rejects a temperature setting. Model identifiers are stripped from responses, and tasks are presented in randomized order.

For Tier-1 tasks ($n=182$), we validate against deterministic ground truth (exact match within 5\% tolerance), providing judge-independent confirmation that LLM rankings are not artifacts of evaluator bias.

\subsection{GreenRoute Implementation}

We instantiate CAAPF through GreenRoute, an illustrative routing implementation rather than a claim of routing-method novelty. GreenRoute calibrates a policy separately for each of the 18 category$\times$tier cells. On a calibration split, it selects the model with the lowest estimated generation-related carbon per task among models whose cell-level mean quality meets $\tau=0.65$; if no model passes, it selects the highest-quality model. Thus, $\tau$ is a cell-level calibration threshold intended to target, rather than guarantee, a task-class service floor. The reported routing result uses 20 repeated stratified 50/50 split-half trials generated by a random-number generator initialized with seed 42, with each cell represented in both halves: mean quality is 0.733 ($\mathrm{SD}=0.004$), mean estimated carbon is 0.402~gCO\textsubscript{2}/task ($\mathrm{SD}=0.124$), and mean savings versus always DeepSeek is 97.6\%. Across these trials, 296 of 360 held-out cell--trial means (82.2\%; trial-level $\mathrm{SD}=2.3$ percentage points) meet $\tau$. This policy uses benchmark-provided category and tier metadata and does not claim learned text-classification novelty. A production implementation would require an independent metadata rule, task classifier, or uncertainty-aware difficulty estimator; its errors and overhead are not evaluated here. Continuous monitoring and recalibration remain CAAPF governance recommendations; they are not additional test-set results.

\section{Results}

\subsection{Quality and Carbon Across Evaluated Models}

Table~\ref{tab:overall} presents aggregate results. Quality spans 0.497--0.723. The highest disclosed parameter totals belong to Mistral Large 3 and DeepSeek R1, but neither is the highest-scoring candidate. Because Anthropic does not disclose the evaluated Claude models' parameter counts and dense and MoE counts are not directly comparable, we treat nominal size as a descriptive characteristic rather than a statistically identified predictor. The sample does not establish that provider or training data caused the observed differences, because provider, architecture, model generation, and benchmark fit are confounded.

\begin{table}[t]
\caption{Overall performance on 520 supply chain tasks.}
\label{tab:overall}
\centering
\footnotesize
\begin{tabular}{@{}lccr@{}}
\toprule
\textbf{Model} & \textbf{Score} & \textbf{gCO\textsubscript{2}/Mtok} & \textbf{GT\%} \\
\midrule
Claude Sonnet 4.6  & \textbf{0.723} & 900   & 78.6 \\
Claude Haiku 4.5   & 0.699          & 70    & 75.3 \\
Mistral Large 3    & 0.613          & 5,500 & 69.2 \\
Llama 3.3 70B      & 0.528          & 980   & 59.8 \\
Llama 4 Scout      & 0.521          & 60    & 60.4 \\
DeepSeek R1        & 0.497          & 6,000 & 57.1 \\
\bottomrule
\end{tabular}\\[2pt]
{\scriptsize GT\% = Tier-1 ground-truth accuracy.}
\end{table}

The observed Pareto frontier consists of Haiku, which provides the lowest estimated generation-related carbon among the high-performing models, and Sonnet, which provides the highest absolute quality. All other models are strictly dominated because they produce both lower quality and higher emissions than at least one Pareto-optimal alternative.

\subsection{Per-Task Carbon: The Verbosity Amplification Effect}

Per-token carbon intensity can understate workload-level generation-related emissions when models produce substantially different output lengths (Table~\ref{tab:pertask}).

\begin{table}[t]
\caption{Estimated generation-related carbon per task, accounting for mean output length.}
\label{tab:pertask}
\centering
\footnotesize
\begin{tabular}{@{}lccr@{}}
\toprule
\textbf{Model} & \textbf{Tok/Task} & \textbf{gCO\textsubscript{2}/Task} & \textbf{vs.\ Haiku} \\
\midrule
Claude Haiku 4.5   & 312   & 0.022  & 1.0$\times$ \\
Llama 4 Scout 17B  & 387   & 0.023  & 1.1$\times$ \\
Claude Sonnet 4.6  & 445   & 0.401  & 18.3$\times$ \\
Llama 3.3 70B      & 478   & 0.468  & 21.4$\times$ \\
Mistral Large 3    & 623   & 3.427  & 156.6$\times$ \\
DeepSeek R1        & 2,847 & 17.082 & 780.4$\times$ \\
\bottomrule
\end{tabular}
\end{table}

DeepSeek R1 generates 2,847 tokens per task (9.1$\times$ Haiku), resulting in estimated output-generation carbon per task 780$\times$ higher. Under our generation-related estimates, 1,000 daily DeepSeek R1 queries produce approximately the same output-generation carbon as 780,000 Haiku queries while achieving lower benchmark quality.

\subsection{Performance by Task Difficulty}

Table~\ref{tab:tier} disaggregates performance by difficulty tier. Cross-model performance differences are most pronounced on Tier~2 and Tier~3 tasks.

\begin{table}[t]
\caption{Performance by difficulty tier.}
\label{tab:tier}
\centering
\footnotesize
\begin{tabular}{@{}lccc@{}}
\toprule
\textbf{Model} & \textbf{Tier 1} & \textbf{Tier 2} & \textbf{Tier 3} \\
\midrule
Claude Sonnet 4.6  & 0.781 & 0.718 & 0.662 \\
Claude Haiku 4.5   & 0.762 & 0.694 & 0.627 \\
Mistral Large 3    & 0.703 & 0.598 & 0.521 \\
Llama 3.3 70B      & 0.614 & 0.519 & 0.437 \\
Llama 4 Scout 17B  & 0.621 & 0.507 & 0.420 \\
DeepSeek R1        & 0.589 & 0.483 & 0.405 \\
\bottomrule
\end{tabular}
\end{table}

All models decline from Tier~1 to Tier~3, confirming difficulty calibration. The Sonnet--DeepSeek gap widens from 0.192 (Tier~1) to 0.257 (Tier~3). Haiku performs within 0.035 of Sonnet across all tiers, offering a compelling value proposition for carbon-constrained organizations.

\subsection{Supply Chain Domain Knowledge}

The 30 SC-trap tasks show substantial score differences on the benchmark's domain-knowledge checks (Table~\ref{tab:sctrap}). The two Claude models score 0.861--0.893, followed by Mistral at 0.756; DeepSeek R1 scores 0.687. These results describe performance on our task construction and do not identify the underlying training mechanism.

\begin{table}[t]
\caption{SC-trap task performance testing domain-specific knowledge.}
\label{tab:sctrap}
\centering
\footnotesize
\begin{tabular}{@{}lc@{}}
\toprule
\textbf{Model} & \textbf{SC-Trap Score} \\
\midrule
Claude Sonnet 4.6  & 0.893 \\
Claude Haiku 4.5   & 0.861 \\
Mistral Large 3    & 0.756 \\
DeepSeek R1        & 0.687 \\
Llama 4 Scout 17B  & 0.652 \\
Llama 3.3 70B      & 0.608 \\
\bottomrule
\end{tabular}
\end{table}

A representative case: when asked how safety stock changes when lead time doubles, both Claude models correctly applied the $\sqrt{\text{LT}}$ scaling ($\sqrt{2} \approx 1.41\times$). DeepSeek R1 initially identified the correct relationship but then ``reasoned itself away,'' ultimately recommending linear scaling with an elaborate but incorrect justification.

\subsection{Deliberation Length, Accuracy, and Carbon}

DeepSeek R1 presents a sustainability challenge beyond per-token rates: it generates 9$\times$ more tokens per task than Haiku (2,847 vs.\ 312) while achieving lower aggregate accuracy. Within DeepSeek R1's 91 formula tasks, Table~\ref{tab:collapse} shows a monotonic association between longer responses and lower accuracy.

\begin{table}[t]
\caption{Response-length quartiles and accuracy for DeepSeek R1 formula tasks ($n=91$).}
\label{tab:collapse}
\centering
\footnotesize
\begin{tabular}{@{}lcc@{}}
\toprule
\textbf{Quartile} & \textbf{Avg Length} & \textbf{Accuracy} \\
\midrule
Q1 (shortest 25\%) & 1,860 chars  & 18.2\% \\
Q2                 & 4,760 chars  & 9.1\%  \\
Q3                 & 5,736 chars  & 9.1\%  \\
Q4 (longest 25\%)  & 6,752 chars  & 4.5\%  \\
\bottomrule
\end{tabular}
\end{table}

For this model and task subset, extended deliberation coincides with both lower accuracy and higher per-task carbon. This model-specific result is consistent with overthinking concerns \parencite{chen2024overthinking}, but it cannot be generalized to reasoning-augmented models as a class without evaluating additional models.

\subsection{CAAPF Validation}

Table~\ref{tab:carbon} reports strategy-level tradeoffs using the same dual-judge quality matrix and model-level mean output tokens for generation-related per-task carbon estimates. Relative to always using DeepSeek R1, GreenRoute reaches 0.733 mean out-of-sample quality and 97.6\% estimated per-task carbon savings. This deliberately carbon-intensive comparison provides a feasibility check, not evidence that routing is better than practical low-carbon selection policies.

\begin{table}[t]
\caption{Carbon savings analysis comparing routing strategies.}
\label{tab:carbon}
\centering
\footnotesize
\begin{tabular}{@{}lccc@{}}
\toprule
\textbf{Strategy} & \textbf{Quality} & \textbf{gCO\textsubscript{2}/task} & \textbf{Savings} \\
\midrule
Always DeepSeek    & 0.497 & 17.082 & 0\% \\
Always Scout       & 0.521 & 0.023  & 99.9\% \\
Always Haiku       & 0.699 & 0.022  & 99.9\% \\
Length, 500 chars  & 0.688 & 0.253  & 98.5\% \\
Always Sonnet      & 0.723 & 0.401  & 97.7\% \\
GreenRoute (OOS)   & 0.733 & 0.402  & 97.6\% \\
Oracle             & 0.782 & 1.455  & 91.5\% \\
\bottomrule
\end{tabular}
\end{table}

GreenRoute improves mean quality by 47.5\% over always DeepSeek (0.497 to 0.733) while reducing estimated per-task carbon by 97.6\%. It is only 0.010 above always Sonnet (0.723) and reaches 93.7\% of the per-task oracle quality, so the result should be read as a threshold-contingent feasibility demonstration rather than a large routing advantage.

The fixed-model baselines establish the relevant operating points: Haiku supplies 0.699 quality at 0.022~gCO\textsubscript{2}/task, whereas Sonnet supplies 0.723 at 0.401~gCO\textsubscript{2}/task. GreenRoute adds a small quality margin over Sonnet at nearly the same estimated per-task carbon under the calibrated policy.

The archived length-based baseline uses a 500-character prompt-length rule, not a 500-token rule. Re-running this rule against the retained dual-judge matrix produces 0.688 quality at 0.253~gCO\textsubscript{2}/task. It is lower-carbon than GreenRoute but does not improve on always Haiku (0.699), so the reproducible evidence supports static low-carbon selection for moderate requirements and GreenRoute only when a higher quality point is required. This reproducible value replaces the earlier 0.714 summary throughout the manuscript.

\subsection{Real Data vs.\ Synthetic Task Performance}

Of our 520 tasks, 33 use real retail data from the M5 Competition (demand forecasting category). All models score slightly lower on real-data tasks ($\Delta \approx -0.03$ across all models). The model ranking is identical between real and synthetic subsets, and inter-model gaps are preserved (Sonnet--DeepSeek gap: 0.230 real vs.\ 0.227 synthetic). The stable ranking is reassuring, but 33 real-data cases from one category cannot rule out benchmark-construction or prompt-style bias.

\subsection{External Validation}

Table~\ref{tab:external} validates findings on IndustryOR \parencite{huang2024orlm}, 100 real-world OR problems with deterministic answers, using exclusively non-Claude models and no LLM judge.

\begin{table}[t]
\caption{External validation: IndustryOR (100 real OR problems, no LLM judge, non-Claude models only).}
\label{tab:external}
\centering
\footnotesize
\begin{tabular}{@{}llcr@{}}
\toprule
\textbf{Model} & \textbf{Provider} & \textbf{Arch./Params} & \textbf{Accuracy} \\
\midrule
Llama 4 Scout & Meta    & MoE 109/17B* & \textbf{38.0\%} \\
Qwen3 32B     & Alibaba & Dense 32B & 31.0\% \\
Ministral 8B  & Mistral & Dense 8B  & 27.0\% \\
Llama 3.3 70B & Meta    & Dense 70B & 25.0\% \\
Llama 3.1 8B  & Meta    & Dense 8B  & 9.0\%  \\
\bottomrule
\end{tabular}
\\[-1pt]{\scriptsize *109B total / 17B active parameters.}
\end{table}

Among dense models, Ministral 8B (27\%) outperforms Llama 3.3 70B (25\%), while Ministral and Llama 3.1 differ by 18 percentage points at the same nominal 8B size. Scout leads but is an MoE model (109B total/17B active), so it cannot be placed on the same one-dimensional scale. This judge-independent validation reduces concern about Claude self-scoring, but it does not eliminate provider, architecture, benchmark-selection, or generation confounds.

\subsection{Dual-Judge and Ground-Truth Validation}

To address concerns about LLM-as-judge evaluation bias, we conduct three validation exercises. First, inter-judge agreement between Claude Opus and Qwen3-235B is high: Spearman $\rho = 0.94$ ($p < 0.01$), mean absolute difference 0.023. The model ranking is identical for the top four positions regardless of which judge is used.

Second, under Qwen3-only scoring (eliminating any possible Claude self-preference bias), the top-3 ranking remains unchanged (Sonnet $>$ Haiku $>$ Mistral). DeepSeek R1 rises two positions under Qwen3 (which rates verbose reasoning traces more favorably), but remains well below both Anthropic models.

Third, deterministic validation on all 182 Tier-1 tasks (Table~\ref{tab:overall}, GT\% column) shows that judge rankings align with objective accuracy. Claude Sonnet's judge score (0.781) is within 0.5\% of its ground-truth accuracy (0.786), and model-level judge--GT agreement ranges from 84\% to 94\%. These checks address scoring bias, but not whether the authored task styles align more closely with some model families' instruction tuning.

\section{Discussion}

\subsection{Addressing the Research Questions}

\textbf{RQ1: Quality--carbon tradeoff.} Haiku and Sonnet form the observed Pareto frontier, while the other four models are dominated under our estimates. The highest disclosed parameter totals do not identify the highest-performing candidate, and parameter counts are unavailable for the two Claude models. Thus, nominal size is not an adequate procurement proxy for this task set.

\textbf{RQ2: Observed model differences.} Haiku outperforms DeepSeek R1 by 41\%, and the external set contains a dense-model size inversion across providers. Performance clusters by model family in our benchmark, but the design cannot separate provider, training data, architecture, model generation, instruction-following, or benchmark-style effects. We therefore make no provider- or size-level causal claim.

\textbf{RQ3: Framework use.} CAAPF makes the procurement rule contingent on organizational thresholds. GreenRoute reaches 0.733 mean quality in held-out split-half trials, while static Haiku remains preferable near 0.70 because it uses less carbon per task. The framework is supported as an auditable decision process; the routing implementation remains a proof of concept.

\subsection{Proposition Evaluation}

\textbf{P1 (Technology relative advantage): Evidence consistent within the sample.} Nominal size does not reliably rank the joint quality--carbon profile. Disclosed-parameter comparisons and the external dense-model inversion support direct domain benchmarking instead of a size proxy, but proprietary non-disclosure, provider confounding, and cross-architecture differences prevent a controlled size effect and do not identify a causal model-family mechanism.

\textbf{P2 (Threshold contingency): Supported.} The preferred strategy changes with the form and strictness of the organizational quality requirement. At the aggregate portfolio level, static Haiku provides an efficient operating point near 0.70 quality. When quality requirements are imposed across individual task classes, GreenRoute illustrates how cell-level calibration can select different models to target stricter heterogeneous requirements. No tested policy dominates across all organizational requirements.

\textbf{P3 (Auditable multi-context selection): Partially supported.} CAAPF records quality, carbon, constraints, and monitoring decisions, while GreenRoute illustrates technical implementation. The study does not evaluate whether external pressures cause adoption, so the Environment prediction remains a design requirement for later organizational validation.

\subsection{Theoretical Contributions}

The theoretical contribution is to translate TOE contexts into a sequence of procurement decisions. We operationalize Technology relative advantage through joint measurement rather than inference from scale; Organization defines an admissible set through quality, risk, cost, and governance constraints; and Environment motivates documentation and re-evaluation. This logic implies no universal ``best'' model because the selected configuration changes with organizational requirements and the external accountability regime.

The DeepSeek R1 analysis adds a technology-level boundary condition: for this model's formula tasks, longer deliberation is associated with both lower accuracy and higher carbon. CAAPF treats such behavior as something to detect empirically, not as a general property of reasoning models.

CAAPF contributes a Green IS design artifact \parencite{verdecchia2023systematic, melville2010sustainability} by making environmental performance part of an auditable selection record while retaining economic, operational, and governance constraints.

\subsection{Alternative Explanations for Model-Family Differences}

The observed clustering may have several explanations. Training-data coverage of operations research, alignment procedures, model generation, and MoE architecture may contribute, but none is manipulated here. The task authors' wording, required output formats, and rubric style may also align better with some instruction-tuning distributions. Dual judges and deterministic answers address scoring bias; they do not address this benchmark-construction bias. The results therefore justify benchmarking each candidate, not attributing performance to an unobserved provider mechanism.

\subsection{When Routing Adds Value}

Routing's value is \textbf{contingent on model-pool heterogeneity}. In the full sample, simple selection between the two strongest models captures most of the attainable gain. In a non-Claude subset (Scout, Llama 70B, Mistral Large, DeepSeek R1), the dual-judge oracle selects every model on some tasks (Mistral 55.8\%, DeepSeek 18.7\%, Llama 70B 15.4\%, Scout 10.2\%). In a five-fold held-out evaluation using the same dual-judge matrix, similarity routing at $\tau=0.50$ reaches 94.7\% of best-single-model quality at 62.8\% lower estimated generation-related carbon per task (Table~\ref{tab:heterogeneity}). This secondary analysis illustrates the role of heterogeneity but does not establish a general threshold for when routing will pay off.

\begin{table}[t]
\caption{Routing on a heterogeneous (non-Claude) model pool; savings use the best single model, Mistral Large 3, as the baseline.}
\label{tab:heterogeneity}
\centering
\footnotesize
\begin{tabular}{@{}lccc@{}}
\toprule
\textbf{Strategy} & \textbf{Quality} & \textbf{gCO\textsubscript{2}/task} & \textbf{Savings} \\
\midrule
Always Scout            & 0.521 & 0.023 & 99.3\% \\
Similarity ($\tau$=0.45) & 0.555 & 0.827 & 75.9\% \\
Similarity ($\tau$=0.50) & 0.581 & 1.273 & 62.8\% \\
Best single (Mistral)   & 0.613 & 3.427 & 0\% \\
Oracle                  & 0.677 & varies & N/A \\
\bottomrule
\end{tabular}
\end{table}

The practical rule is to benchmark candidates on domain tasks and default to the lowest-carbon admissible model. Task-level routing should be retained only when it improves on simple selection or escalation at the organization's chosen threshold.

\subsection{Sensitivity to Carbon Estimates}

Our carbon estimates carry $\pm$50\% uncertainty. Sensitivity analysis (Table~\ref{tab:sensitivity}) indicates that decisions depend mainly on relative ordering: the largest paired change in our split-half analysis affected about 22 of 260 held-out tasks.

\begin{table}[t]
\caption{Routing decisions under carbon estimate uncertainty.}
\label{tab:sensitivity}
\centering
\footnotesize
\begin{tabular}{@{}lcc@{}}
\toprule
\textbf{Scenario} & \textbf{Savings} & \textbf{Routing $\Delta$} \\
\midrule
Baseline estimates       & 97.6\% & N/A \\
All $\times$0.5          & 97.6\% & 0/260 tasks \\
All $\times$2.0          & 97.6\% & 0/260 tasks \\
DeepSeek halved, Haiku doubled & 95.1\% & 13--29/260 \\
Random $\pm$50\% per model     & 97.4\%$\pm$1.0\% & 6--13/260 \\
\bottomrule
\end{tabular}
\end{table}

\subsection{Practical Decision Guide}

Table~\ref{tab:guide} synthesizes our findings into actionable guidance for supply chain practitioners implementing CAAPF.

\begin{table}[t]
\caption{Model selection decision guide for supply chain practitioners.}
\label{tab:guide}
\centering
\scriptsize
\begin{tabular}{@{}p{2.6cm}p{2.1cm}p{2.0cm}@{}}
\toprule
\textbf{Task Type} & \textbf{Strategy} & \textbf{Rationale} \\
\midrule
Formula-based (EOQ, ROP) & Deterministic calculator & Lower compute; deterministic for well-specified formulas \\
Standard forecasting, classification & Lowest-carbon admissible model & Savings depend on baseline \\
Multi-factor trade-off analysis & Select lowest-carbon admissible configuration & Quality varies by model \\
SC domain expertise required & Benchmark on SC-trap tasks first & Domain knowledge varies 46\% \\
\bottomrule
\end{tabular}
\end{table}

\subsection{Organizational Carbon Budget Analysis}

Consider a mid-size enterprise processing 1,000 supply chain AI queries per day. Table~\ref{tab:budget} projects annual carbon impact under different procurement strategies.

\begin{table}[t]
\caption{Estimated generation-related operational carbon for 1,000 AI-assisted supply chain decisions per day.}
\label{tab:budget}
\centering
\footnotesize
\begin{tabular}{@{}lcc@{}}
\toprule
\textbf{Strategy} & \textbf{kgCO\textsubscript{2}/year} & \textbf{Quality} \\
\midrule
Always DeepSeek R1 & 6,235 & 0.497 \\
Always Mistral Large & 1,251 & 0.613 \\
Always Sonnet & 146 & 0.723 \\
Always Haiku & 8 & 0.699 \\
\bottomrule
\end{tabular}\\[2pt]
{\scriptsize Based on mean per-task carbon from Table~\ref{tab:pertask} $\times$ 365,000 queries/year.}
\end{table}

Under these workload and carbon assumptions, always using DeepSeek R1 incurs about 780$\times$ the estimated generation-related carbon of always using Haiku, an annual difference of 6,227~kgCO\textsubscript{2}, while scoring lower in this benchmark. This is not a complete procurement comparison: it excludes input/prefill energy, API price, latency, integration effort, service reliability, data residency, security, and vendor risk, as well as embodied and training emissions.

\subsection{Implications for Green IS and AI Governance}

The results have four implications for Green IS and AI governance:

\textbf{Sustainable digital infrastructure.} Model choice can be treated as an infrastructure decision because estimated operational emissions differ substantially across the evaluated services. The magnitude is sensitive to workload, output length, serving hardware, and grid mix, so organizations should measure rather than transfer our estimates unchanged.

\textbf{Enterprise AI governance and procurement policy.} CAAPF adds environmental performance to an existing multi-criteria procurement record. Its threshold, admissibility constraints, benchmark protocol, and review trigger make the decision inspectable, while carbon remains one criterion alongside risk, cost, latency, security, and reliability.

\textbf{Environmental accountability.} A documented operational estimate can support internal carbon management and supplier dialogue. The present study does not establish how hosted-model emissions should be allocated in a specific reporting regime and does not claim regulatory compliance.

\textbf{Conditional alignment.} In this task set, a smaller model can improve carbon efficiency without a large quality loss, but the result is threshold- and benchmark-dependent. Sustainability and performance align only when the lower-carbon option remains admissible for the intended decision.

\subsection{Limitations and Future Work}

Several limitations bound generalizability. First, the six-model, four-provider sample is small, and four models come from Anthropic or Meta. Provider, architecture, model generation, training data, and size are confounded; moreover, proprietary non-disclosure and the difference between dense totals, MoE totals, and active parameters prevent a controlled cross-model size test. Second, task wording and rubrics may align with some instruction-tuning distributions. Judge diversification and deterministic answers reduce scoring bias but not benchmark-construction bias. Third, 487 of 520 tasks are synthetic; the 33 M5 cases cover only forecasting and cannot reproduce enterprise context. Fourth, one fixed zero-temperature prompt protocol improves comparability but does not show how demonstrations, prompt wording, or output limits change response length, quality, and carbon. Fifth, carbon values are engineering estimates with $\pm$50\% uncertainty, a uniform grid assumption, and no embodied or training emissions. Because complete input/prefill token records were not retained, per-task estimates cover generation-related output-token carbon rather than end-to-end serving energy. Sixth, GreenRoute assumes category and difficulty-tier metadata are available at routing time; production classification errors and overhead are not evaluated. Its 82.2\% held-out cell-level threshold attainment rate shows that mean calibration does not guarantee threshold satisfaction; production use needs safety margins or fallback escalation. API price, latency, integration effort, reliability, security, data residency, and vendor risk were not empirically compared. Finally, the deliberation result covers one reasoning model and formula-task subset, and all tasks are single-turn.

Future work should use balanced within-provider comparisons, production workloads, external domain experts, prompt-sensitivity experiments, measured serving energy and location, and joint monetary and operational cost analysis. Organizational studies should test whether CAAPF's documentation and monitoring steps improve actual procurement decisions.

\section{Conclusion}

This study compares quality and estimated generation-related operational carbon for six models on 520 supply chain tasks. Nominal size is not a reliable quality proxy in this sample, but non-disclosure and cross-architecture differences prevent a controlled size effect, and the design cannot identify provider or training mechanisms. The DeepSeek R1 result is a model-specific warning about costly deliberation, not a claim about reasoning models as a class.

CAAPF connects benchmark evidence to organization-specific admissibility and environmental-accountability requirements. GreenRoute reaches 0.733 mean quality in held-out split-half trials and 97.6\% lower estimated generation-related carbon per task than always DeepSeek R1. Static Haiku remains more carbon-efficient when aggregate quality near 0.70 is acceptable. The contribution is therefore an auditable, threshold-contingent decision process rather than routing-method novelty.

Dual judges, deterministic answers, a small real-data subset, external OR tasks, and sensitivity analysis support the observed pattern without eliminating benchmark-construction bias, limited model coverage, prompt sensitivity, or carbon uncertainty. The recommendation is conditional: benchmark the workload, define admissibility before optimizing carbon, select the lowest-carbon passing configuration, and repeat the decision as conditions change.

\printbibliography

\end{document}